\documentclass[aps,prl,twocolumn,groupedaddress]{revtex4}  % for review and submission
\usepackage{graphicx}  % needed for figures
\usepackage{dcolumn}   % needed for some tables
\usepackage{bm}        % for math
\usepackage{amssymb}   % for math
\usepackage{float}
\usepackage{color}

\begin{document}

\title{Ultrafast Magneto-Acoustics in Nickel Films}

\widetext

\author{Ji-Wan Kim}
\author{Mircea Vomir}
\author{Jean-Yves Bigot}

\email{bigot@ipcms.u-strasbg.fr}

\affiliation{Institut de Physique et Chimie des Mat\'eriaux de Strasbourg, UMR 7504, CNRS,
Universit\'e de Strasbourg, BP 43, 23 rue du Loess, 67034 Strasbourg Cedex 02, France.}

%\date{\today}

\begin{abstract}
We report about the existence of magneto-acoustic pulses propagating in a 200-nm-thick ferromagnetic nickel film excited with 120 fs laser pulses. They result from the coupling between the magnetization of the ferromagnetic film and the longitudinal acoustic waves associated to the propagation of the lattice deformation induced by the femtosecond laser pulses. The magneto-acoustic pulses are detected from both the front and back sides of the film, using the time-resolved magneto-optical Kerr technique, measuring both the time dependent rotation and ellipticity. We show that the propagating acoustic pulse couples efficiently to the magnetization and is strong enough to induce a precession of the magnetization. It is due to a transient change of the crystalline anisotropy associated to the lattice deformation. It is shown that the results can be interpreted by combining the concepts of acoustic pulse propagation and ultrafast magnetization dynamics.
\end{abstract}

%\pacs{72.55.+s, 75.78.Jp, 78.20.hc, 78.20.Ls}

\maketitle

The technology of information and communication constantly needs to improve the speed and density of memory devices. Towards that goal, intense researches are being carried out for manipulating the spins in magnetic materials using various excitation methods like the use of external magnetic field pulses \cite{Hiebert1997, Acremann2000, Gerrits2002}. Alternatively, femtosecond laser pulses have been utilized to induce an ultrafast demagnetization via a sudden and abrupt change of the temperature of the magnetic material \cite{Beaurepaire1996, Hohlfeld1997, Aeschlimann1997, Scholl1997, Koopmans2000, Guidoni2002, Kampen2002}, or using the inverse Faraday effect \cite{Kimel2005, Hansteen2005}. This new field of magnetism, named "femtomagnetism" \cite{Zhang2002, Bovensiepen2009}, uses photons to directly manipulate magnetic structures with a temporal resolution of a few femtosecond. The demagnetization can then be used to modify the anisotropy of the magnetic material \cite{Bigot2005} which leads to a reorientation of the magnetization vector followed by its precession and damping in the direction opposite to the initial one \cite{Ju1999, Kampen2002, Vomir2005}. In spite of their versatility, due to the various laser wavelengths and pulse durations, magneto-optical methods are limited by the absorption depth of photons. For application purposes, it is a disadvantage for controlling devices at long distances, particularly in opaque materials like ferromagnetic metals.

In the present work we explore an alternative way of controlling the magnetization, based on magneto-acoustic performed at room temperature in ferromagnetic films. It is known that strain pulses, corresponding to a lattice deformation of a material, can be generated with a laser pulse, a subject which has been extensively studied theoretically and experimentally \cite{Grahn1989, Wright1995, Wright2008}, since the pioneering works of Thomsen \textit{et al}. \cite{Thomsen1984, Thomsen1986}. The relative amplitude of these strain pulses can be as large as 10$^{-3}$. Such acoustic waves have been used for perturbing the magnetic properties of a dilute magnetic semiconductor material, at low temperature and with a very low efficiency \cite{Scherbakov2010}. Here we show that one can efficiently use the strain pulses that propagates over long distances in ferromagnetic metals at room temperature and induce very large changes of the magnetization. In addition, to explain our results we propose a model that takes into account the spatio-temporal propagation of the strain pulses coupled to the magnetic anisotropy. Our results open the way to a large number of applications for the ultrafast control of information in magneto-acoustic devices addressed by ultrashort laser pulses.

The experiment consisted in generating longitudinal acoustic waves in nickel films with femtosecond pump pulses and probing the magnetization dynamics on both faces of the film, using the time resolved magneto-optical Kerr effect. The probe pulses $p_{front}$ and $p_{back}$, with a duration of 120 fs and a wavelength 794 nm, are issued from an amplified Titanium Sapphire laser operated at 5 kHz. The pump pulses (150 fs, 397 nm), obtained by second harmonic generation, excite only the front face of the sample with a maximum energy density of 5 mJcm$^{-2}$. They are focused onto the sample within a spot diameter (200 $\mu$m) twice larger than the two probe beams. The reflectivity $R(t)$, polarization rotation $\theta(t)$ and ellipticity $\varepsilon(t)$ of both $p_{front}$ and $p_{back}$ are detected with polarization bridges as a function of the pump and probe delay $t$. We used two poly-crystalline 200-nm-thick Ni films, deposited either on a glass or a sapphire substrate by electron-beam evaporation.

\begin{figure}[t]
\includegraphics [height=9cm,width=8.6cm]{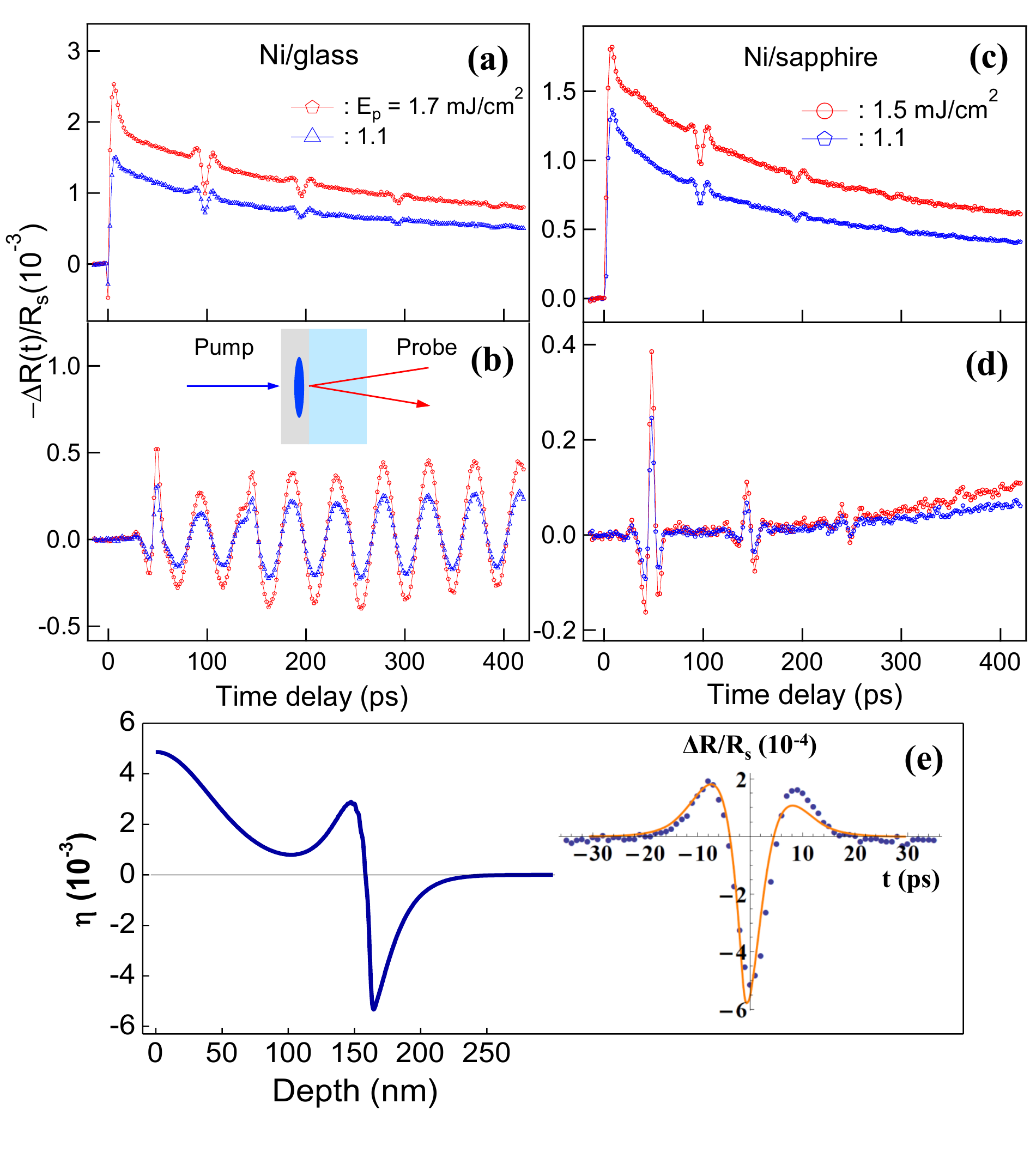}
\caption{\label{PreFig1}Differential reflectivity $\Delta R(t)/R_s$ for different pump energy densities $E_{p}$ of 200-nm-thick nickel films. (a) and (b): Ni/glass sample and (c) and (d): Ni/sapphire sample. (a) and (c) are obtained for $p_{front}$ probing and (b) and (d) for $p_{back}$ probing. (e - left): modeling of the strain profile at t = 40 ps after excitation with a femtosecond laser pulse. (e - right): fit of the differential reflectivity using the model.}
\end{figure}

We first studied the dynamics of the reflectivity for comparison with existing results \cite{Thomsen1986, Saito2003}. Figure 1 shows the differential reflectivity $\Delta R(t)/R_s$ for the Ni/glass film (Figs. 1(a)-1(b)) and Ni/sapphire film (Figs. 1(c)-1(d)) probing the front (Figs. 1(a)-1(c)) and back (Figs. 1(b)-1(d)) sides of the films. Two different pump energy densities $E_{p}$ have been used as shown in the figures. For $p_{front}$, in addition to the well known thermal dynamics associated to a heating of the electrons and lattice, four acoustic echoes with a period of  $T$ = 98 ps are observed. For $p_{back}$ the dynamics of $\Delta R(t)/R_s$ differs for the two substrates with oscillations and echoes in the case of Ni/glass and echoes only in the case of Ni/sapphire. We model $\Delta R(t)/R_s$ by taking into account the dynamical changes of the thermo-optic and piezo-optic properties affecting the complex refractive index $\tilde{n}$ as follows:
\begin{eqnarray}
\Delta \tilde{n} & = & \frac{\partial n}{\partial T} \Delta T_e + i \frac{\partial \kappa}{\partial T} \Delta T_e + \frac{\partial n}{\partial \eta} \eta + i \frac{\partial \kappa}{\partial \eta} \eta
\end{eqnarray}
\noindent $n$ and $\kappa$ are the real and imaginary parts of $\tilde{n}$ at the wavelength of the probe beam, $T_e$ is the electron temperature and $\eta$ is the longitudinal strain pulse. The partial derivatives are the thermo-optic and piezo-optic coefficients. The thermo-optic part essentially involves the electron temperature change \cite{Sun1994}, while $\eta$ and the piezo-optic part occur via the temporal gradient of the lattice temperature $T_l (z,t)$ \cite{Wright1995}. Both temperatures are coupled and can be modeled by the two temperatures model:
\begin{eqnarray}
C_{e}\left(T_{e}\right) \frac{\partial T_e}{\partial t} & = & \ \frac{\partial}{\partial z}(\sigma \frac{\partial T_e}{\partial z})-g \left(T_{e}-T_{l} \right) + \frac{1}{\zeta} P(t) e^{-\frac{z}{\zeta}}\nonumber\\
C_{l}\left(T_{l}\right) \frac{\partial T_{l}}{\partial t} & = & \ g \left(T_{e}-T_{l} \right)
\end{eqnarray}
\noindent where $C_e (C_l)$ is the electron (lattice) heat capacity per unit volume, $\sigma$ the thermal conductivity, $g$ the electron-lattice coupling constant, $\zeta$ the absorption length, $P(t)$ the absorbed pump energy and $z$ is the direction of propagation perpendicular to the film surface. For nickel we used the following values: $\gamma = 6 \times 10^3 $ Jm$^{-3}$K$^{-2}$ ($C_e =\gamma T_e)$ \cite{Beaurepaire1996}, $\sigma = 91$ Wm$^{-1}$K$^{-1}$ \cite{Saito2003}, $\zeta$ = 13.5 nm for $\tilde{n}'$ = 1.61 + 2.36i at 400 nm \cite{ConstSolids1986}. $C_l$ and $g$ are fitted from the experimental shape of the echo as shown on the right side of Fig. 1(e). We first compute $T_e (z,t)$ and $T_l (z,t)$ assuming a semi-infinite slab together with the proper boundary conditions and a pump energy density $E_{p}$ = 1.7 mJ/cm$^2$. Then $T_l (z,t)$ acts as a source term in the following one dimensional wave equation \cite{Wright1995}.
\begin{eqnarray}
\rho \frac{\partial^2 u(z,t)}{\partial t^2}=\rho v^2 \frac{\partial^2 u(z,t)}{\partial z^2}-3\beta B \frac{\partial \delta T_l}{\partial z}
\end{eqnarray}
\noindent $u$ ($\eta=\partial u / \partial z$) is the lattice displacement, $\rho$ is the mass density $(8.91\times10^3$ kgm$^{-3}$), $v$ is the sound velocity in Ni ($4.08\times10^3$ ms$^{-1}$, as measured here), $\beta$ is the linear thermal expansion coefficient $(1.3\times10^{-5}$ K$^{-1}$), $B$ is the bulk modulus ($1.8\times10^{11}$ Nm$^{-2}$ \cite{Luo2011}) of nickel. The strain profile obtained with Eq. 3 for $t =$ 40 ps is shown in Fig. 1(e) (left curve). It consists of a quasi stationary part near the surface of the film and a propagating part which is partially transmitted to the substrate and partially reflected back to the film with a reflection coefficient $r_{ac} \sim -0.39$, which is obtained from successive echoes in Fig. 1(a). The negative sign takes into account the reversal of the strain pulses at interfaces. The propagation of the strain, back and forth in the film, then gives rise to the series of echoes observed in Fig. 1(a)-1(c) with a period $T=2d/v$. The corresponding reflectivity can be obtained after considering the piezo-optic effect in whole depth using \cite{Saito2003}:
\begin{eqnarray}
\frac{\delta r}{r}=-i\frac{4 \pi}{\lambda}\delta z + i \frac{8 \pi \tilde{n}}{\lambda (1-\tilde{n}^2)} \frac{\partial \tilde{n}}{\partial \eta} \int^{\infty}_{0} \eta(z,t) \text{exp}(i \frac{4 \pi \tilde{n}}{\lambda}z)dz \nonumber
\end{eqnarray}
\noindent with $\tilde{n} ({\lambda} = 800$ nm) = 2.48 + 4.38i \cite{ConstSolids1986} and $\delta r/r$ is the differential reflectance ($\Delta R(t)/R_s = 2 Re[\delta r/r]$). The contribution of the strain pulse on the reflectivity is compared with the experimental data on the right side of Fig. 1(e). The fitting parameters are: $C_l = 2.0 \times10^6$ Jm$^{-3}$K$^{-1}$ and $g = 5\times10^{17}$ Wm$^{-3}$K$^{-1}$.

To distinguish the piezo-optic contribution from the thermo-optic one on $\Delta R(t)/R_s$ and the magnetic properties described hereafter we analyze the dynamics on the back side of the samples using $p_{back}$. As seen in Figures 1(b) and 1(d) the echoes now appear at $(2m+1)T/2, (m=0, 1, 2)$ with no thermal component like in the case of $p_{front}$. In addition, for Ni/glass large oscillations occur which are due to interferences between the probe beam reflected at the Ni-glass interface and secondary beams reflected at the strain pulse transmitted partially into the substrate \cite{Thomsen1986,Lin1991}. This is a known effect for transparent materials \cite{Bosco2002}, and is not due to the reflectivity change of Ni. Since the sapphire substrate has a negligible piezo-optic coupling near 800 nm \cite{Wright2008}, these interferences do not show up in Fig. 1(d). Let us notice that, while the strain pulse propagates back and forth in the film, the tensile part of the pulse comes first up to the film surface, the compressive part, on the contrary, arrives first at Ni-glass interface. This reverses the polarity of the differential reflectivity by a strain pulse as shown in Fig. 1(a) and (b).

Let us now focus on the ultrafast magneto-acoustic properties of Ni. In order to investigate the effect of the coupling between the strain pulse and the magnetization, we measured the differential magneto-optical Kerr response $\Delta \theta / \theta_s$ and $\Delta \varepsilon / \varepsilon_s$. Figure 2(a) shows $\Delta \theta / \theta_s$ for different angles of the magnetic field $\phi = 25, 35$, and $45^{\circ}$, with respect to the normal to the sample, for $E_{p}$ = 2.0 mJcm$^{-2}$. Each graph is shifted for clarity and $\Delta R(t)/R_s$ is shown for comparison. The oscillations on the rotation $\theta$ correspond to a precession of the magnetization induced by the laser pulse. Its period decreases with $\phi$ as the effective field increases \cite{Bigot2005}. In addition, two weak magneto-acoustic pulses show up at the same delays $T$ and $2T$ as those on the reflectivity. They are much more contrasted on the ellipticity curves as seen in Figure 2(b) where they are observed up to $t = 3T$.

\begin{figure}[t]
\includegraphics[height=7cm,width=6cm]{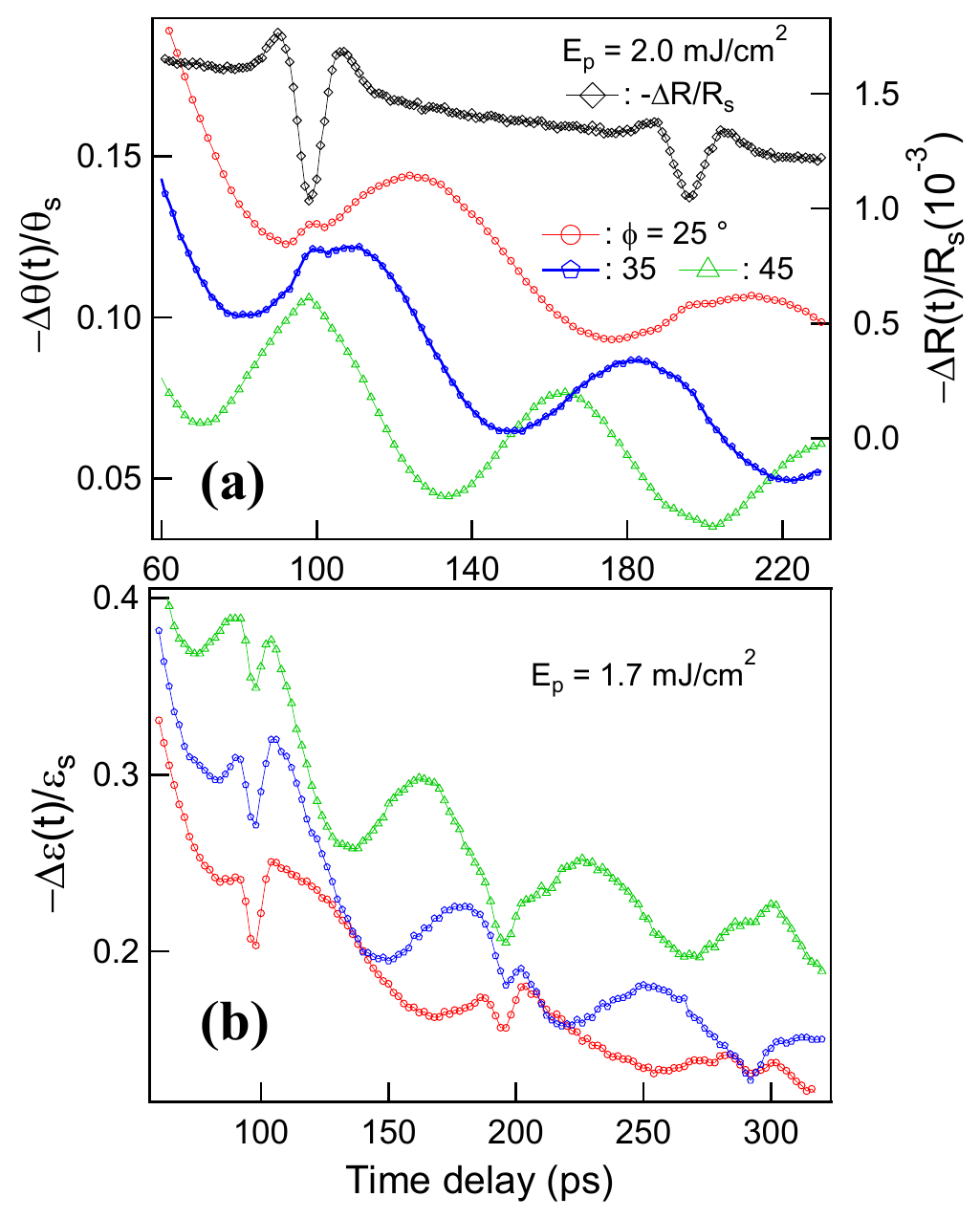}
\caption{\label{PreFig2} Differential Kerr rotation $\Delta \theta / \theta_s$ (a) and ellipticity $\Delta \varepsilon / \varepsilon_s$ (b) probed from the front side of the Ni/glass sample. The curves corresponding to the different angles of the magnetic field $\phi = 25, 35,$ and $45^{\circ}$ have been shifted along the ordinate axis for clarity.}
\end{figure}

In Figure 2, the precession of the magnetization is mainly induced by the thermal effect due to absorption of the femtosecond laser pulse on the front side of the sample as it is already known in several ferromagnetic materials \cite{Ju1999, Kampen2002, Vomir2005}. In addition, the acoustic strain induces a modification of the magnetic anisotropy of nickel through the magneto-elastic property which also contributes to the motion of precession. To separate these two contributions (thermal and strain), we measured the Kerr magneto-optical rotation and ellipticity from the back surface of the sample, using $p_{back}$ as shown in Figure 3. Figure 3(a) displays $\Delta \theta / \theta_s$ for $\phi = 0, 25$, and $35^{\circ}$ with $E_{p}$ = 1.7 mJcm$^{-2}$. Similarly to the reflectivity echoes (Fig. 1(b)), several echoes of magneto-acoustic pulses occur at $(2m+1)T/2, (m=0,1,2)$. They are also much better contrasted on the ellipticity $\varepsilon$ as seen in Figure 3(b). In addition let us stress that, like the polarity reversal of $\Delta R/R_s$ between the front and back side in Fig. 1, the polarity reversals of the magneto-optical responses $\theta$ and $\varepsilon$ also occur when comparing the front and back sides (Figures 2 and 3).

\begin{figure}[t]
\includegraphics[height=7cm,width=6cm]{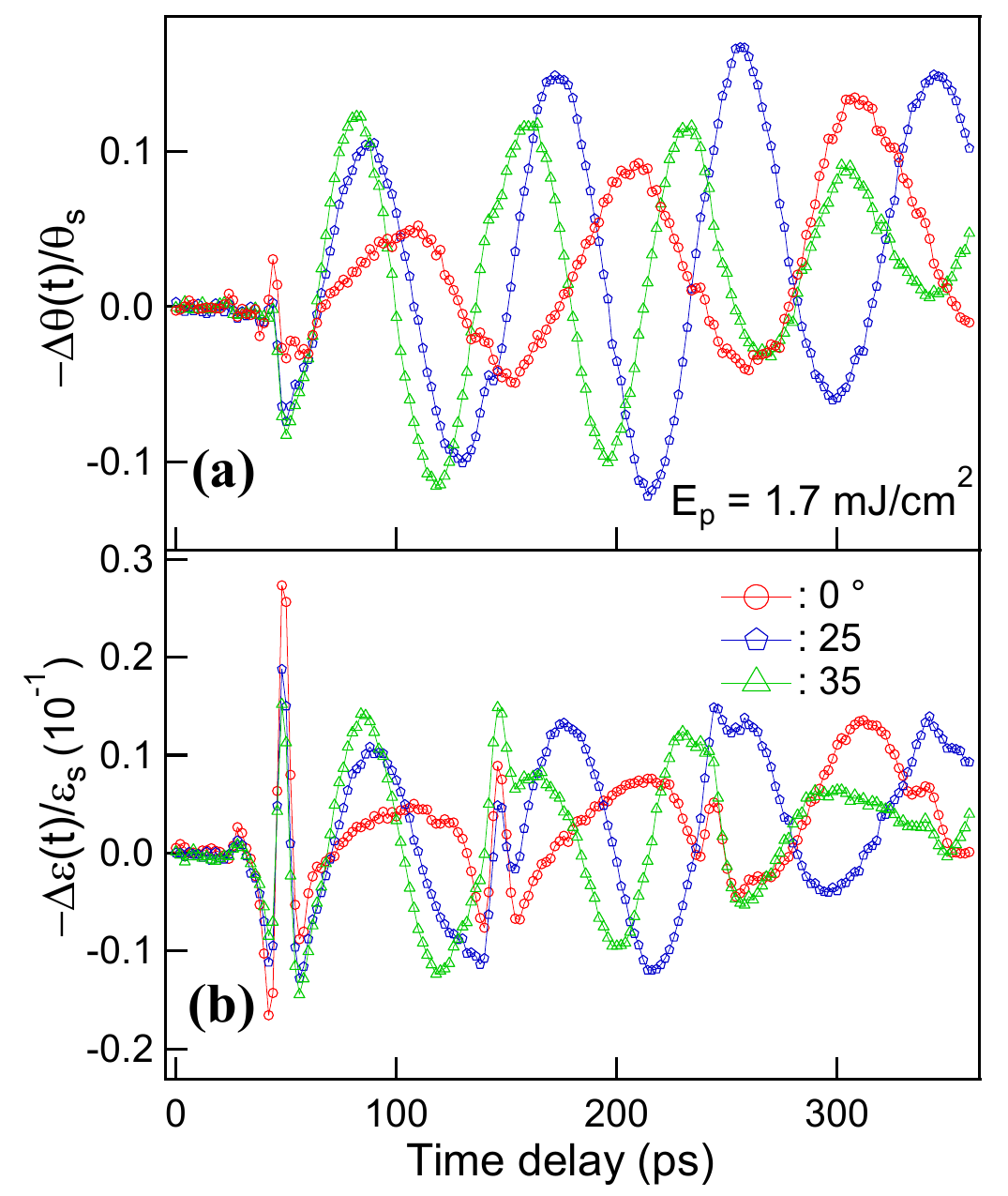}
\caption{\label{PreFig3} Differential Kerr rotation $\Delta \theta / \theta_s$ (a) and ellipticity $\Delta \varepsilon / \varepsilon_s$ (b) probed from the back side of the Ni/glass sample. The curves corresponding to the different angles of the magnetic field $\phi = 0, 25,$ and $35^{\circ}$ have been shifted along the ordinate axis.}
\end{figure}

We estimate the effect of the strain pulse on the magnetization by using the magneto-elastic energy equation: $E_{me} = -3/2 \lambda_s \sigma_s$ cos$^2 \varphi$, where $\lambda_s = -3.3\times10^{-5}$ is the magnetostriction coefficient of a poly-crystalline Ni film \cite{Chikazumi1997}, $\sigma_s = 3(1- \nu)/(1+ \nu) B \eta$ is the stress, $\nu$ the Poisson's ratio, and $\varphi$ is the angle between the strain direction and the magnetization vector. Our experimental value of $4\times10^{-3}$ for the strain amplitude gives an important change of the effective field direction ($\sim 4^{\circ}$). More importantly, the coupling of the strain to the magnetization induces a large motion of precession with 0.1 contrast ratio on the differential rotation $\Delta \theta / \theta_s$. The detailed modeling of the magneto-acoustic pulses can be performed by adding a third temperature in Eq. 2, corresponding to the spins bath \cite{Bigot2001}, together with the Landau-Lifschitz equation taking into account the non-conservation of the magnetization modulus \cite{Bigot2005} and the time dependent magneto-elastic anisotropy. Although straightforward, this calculation is beyond the scope of the present article.

Our results offer interesting perspectives for applications to ferromagnetic metals. While the magnetization dynamics induced by thermal processes are confined at the surface of the film within the penetration depth of light and heat diffusion length, the  acoustic strain pulse propagates with little dispersion over large distances and can be used to manipulate the effective field $H_{eff}(t)$ of the material at long distances as obviously demonstrated in Figure 3. In turn this dynamical change of $H_{eff}(t)$, which is known to modify the magnetization precession and damping \cite{Kim2011, Bigot2005}, can be used to induce and manipulate a torque on the back side of the sample. Another application relies on the fact that the temporal width and frequency bandwidth of the acoustic pulse can be monitored by tailoring the film thickness \cite{Chen2007}. Therefore a Ni film can be utilized not only as a high frequency dispersion-less generator transducer \cite{Wright1995, Saito2003, Wright1994}, but also as a sensitive receiver transducer based on the magneto-optical detection through the efficient ultrafast magneto-acoustic coupling demonstrated here.

In conclusion, we have shown the existence of ultrashort magneto-acoustic pulses in a nickel film excited with femtosecond laser pulses. We have studied their dynamics and propagation to the backside of the film showing that the large strain deformation changes the magneto-elastic coefficients and induces a torque that makes the magnetization precess. Our results open a new way of controlling magnetic devices at room temperature with a large efficiency.

\begin{acknowledgments}
The authors acknowledge the financial support of the European Research Council with the ERC Advanced Grant ATOMAG (ERC-2009-AdG-20090325 247452).
\end{acknowledgments}

\end{document}